\documentclass[%
 aip,
 amsmath,amssymb,
 longbibliography,
 reprint,%
]{revtex4-1}

\usepackage{graphicx}
\usepackage{dcolumn}
\usepackage{bm}

\usepackage[utf8]{inputenc}
\usepackage[T1]{fontenc}
\usepackage{mathptmx}
\usepackage{xcolor}
\usepackage{soul}
\usepackage{color}
\usepackage{multirow}\usepackage{xspace}
\usepackage{amsmath}
\usepackage{amsfonts}
\usepackage{amssymb}
\usepackage{amsthm,bm}
\usepackage{commath}
\usepackage{setspace}

\usepackage{physics,mathtools}

\usepackage[table]{xcolor}
\usepackage{booktabs}
\usepackage{array}
\usepackage{colortbl}
\usepackage{threeparttable}
\usepackage{xr}
\usepackage{titlesec}

\titleformat{\paragraph}[runin]
  {\normalfont\normalsize\bfseries\itshape}{(\theparagraph)}{1em}{}
\titlespacing*{\paragraph}
  {0pt}{*}{0.5em}

\titleformat{\subparagraph}[runin]
  {\normalfont\normalsize\itshape}{(\thesubparagraph)}{1em}{}
\titlespacing*{\subparagraph}
  {0pt}{*}{0.5em}

\renewcommand\theparagraph{\roman{paragraph}}
\renewcommand\thesubparagraph{\alph{subparagraph}}

\DeclareMathOperator{\erfc}{erfc}
\DeclareMathOperator{\erfcx}{erfcx}

\makeatother

\newcommand{\bz}{{\bf z}}

\newcommand{\bd}{{\bf d}}

\newcommand{\br}{{\bf r}}

\newcommand{\bs}{{\bf s}}

\newcommand{\calD}{\mathcal{D}}

\newcommand{\bsigma}{\mbox{\boldmath $\sigma$}}

\newcommand{\brho}{\mbox{\boldmath $\rho$}}

\newcommand{\be}{\begin{equation}}
\newcommand{\ee}{\end{equation}}
\newcommand{\bea}{\begin{eqnarray}}
\newcommand{\eea}{\end{eqnarray}}
\newcommand{\beqa}{\begin{eqnarray*}}
\newcommand{\eeqa}{\end{eqnarray*}}
\newcommand{\nn}{\nonumber}

\newcommand{\ba}{\begin{array}{c}}
\newcommand{\baa}{\begin{array}{cc}}
\newcommand{\baaa}{\begin{array}{ccc}}
\newcommand{\baaaa}{\begin{array}{cccc}}
\newcommand{\ea}{\end{array}}

\newcommand{\bma}{\left[\begin{array}{c}}
\newcommand{\bmaa}{\left[\begin{array}{cc}}
\newcommand{\bmaaa}{\left[\begin{array}{ccc}}
\newcommand{\bmaaaa}{\left[\begin{array}{cccc}}
\newcommand{\ema}{\end{array}\right]}

\begin{document}

\preprint{AIP/123-QED}

\title{Universal tuning of quantum electrodynamic interactions from power laws to exponential screening and logarithmic antiscreening}
\author{Michael N. Leuenberger}
\affiliation{NanoScience Technology Center, Department of Physics, College of Optics and Photonics, University of Central Florida, Orlando, FL 32826, USA.}
\email{michael.leuenberger@ucf.edu}
\author{Daniel Gunlycke}
\affiliation{U.S. Naval Research Laboratory, 4555 Overlook Ave SW, Washington, DC 20375, USA}
\email{lennart.d.gunlycke.civ@us.navy.mil}



\begin{abstract}
We introduce a material-agnostic platform for \emph{universal tuning of quantum electrodynamic interactions
from power laws to exponential screening and logarithmic antiscreening}, realized in a dielectric spacer bounded by
two gate-tunable two-dimensional conductors. The structured electromagnetic environment is completely
specified by the transverse-magnetic and transverse-electric reflection amplitudes
\(r_{\mathrm{TM/TE}}(q_\perp,\omega)\) of the sheets. Starting from the QED action and a Green-function
formulation, we resum the multiple-reflection series and show that the interactions are governed
by a discrete set of transverse cavity harmonics. In the transparent limit \(r_{\rm TM}\to 0\), the interactions
reduce to bulk power laws \(U(\rho)\propto \rho^{-\alpha}\). In the reflective limit \(|r_{\rm TM}|\to 1\),
the \emph{phase/parity} of \(r_{\rm TM}\) selects two qualitatively distinct branches: a Dirichlet/PEC
(screening) branch \(r_{\rm TM}\to -1\) that removes the gapless transverse mode and yields an evanescent
Bessel-\(K\) function \(U(\rho)\propto e^{-\pi\rho/d}/\sqrt{\rho/d}\) at \(\rho\gg d\), and an opposite
Neumann/PMC-like (antiscreening) branch \(r_{\rm TM}\to +1\) that retains a gapless mode and can strongly
enhance the long-range tail. Thus, the same heterostructure provides in situ electrical control over
both the \emph{range} and the \emph{strength} of mediated interactions.

We highlight a flagship consequence for scalable quantum hardware: an electrically switchable and
range-programmable spin--spin interaction mediated by quantum electrodynamic dipole spin resonance (QED-DSR). The same
tunable Feynman propagator that shapes Coulomb, dipolar, and fluctuation-induced vdW/CP interactions
also controls the QED-DSR exchange, enabling two-qubit couplers with tunable interaction length set by \(d\)
and tunable amplitude and parity set by \(r_{\mathrm{TM/TE}}\). Our results establish tunable
2D conductor--dielectric--conductor heterostructures as a broadly applicable route to reconfigurable
interaction landscapes and electrically programmable spin-qubit couplers.
\end{abstract}


\maketitle


Long--range electromagnetic interactions between localized objects, such as charges, dipoles and spins, are
a primary control knob for collective quantum phases and for engineered quantum devices. In two
dimensions, where screening, confinement and nonlocal response are unusually strong, the ability to
reshape the \emph{functional form} and \emph{range} of interactions can reprogram phase diagrams and can
enable reconfigurable couplers for quantum information hardware. A central challenge is that in most
platforms the interaction is essentially fixed: dielectric environments renormalize amplitudes
but rarely provide an \emph{in situ} dial that continuously interpolates between distinct asymptotic
laws.

Recent progress in van der Waals heterostructures and cavity-enabled materials platforms has made this
opportunity concrete. Two-dimensional conductors with gate-tunable optical response are now routinely
integrated into layered stacks, and intrinsic cavity electrodynamics can emerge in such devices
\cite{Kipp2025Cavity}. In parallel, long-distance spin--spin couplers mediated by electromagnetic
environments have been pursued in cavity and circuit architectures
\cite{Imamoglu1999,BurkardImamoglu2006,Samkharadze2018,Borjans2020,HarveyCollard2022,Dijkema2025}.
These developments motivate a unified question: can one engineer a \emph{universal gate knob} that
continuously switches condensed-matter interactions \emph{from bulk power laws to exponential
screening and logarithmic antiscreening} within a single, broadly applicable device geometry?

\begin{figure*}[t]
\centering
\includegraphics[width=\textwidth]{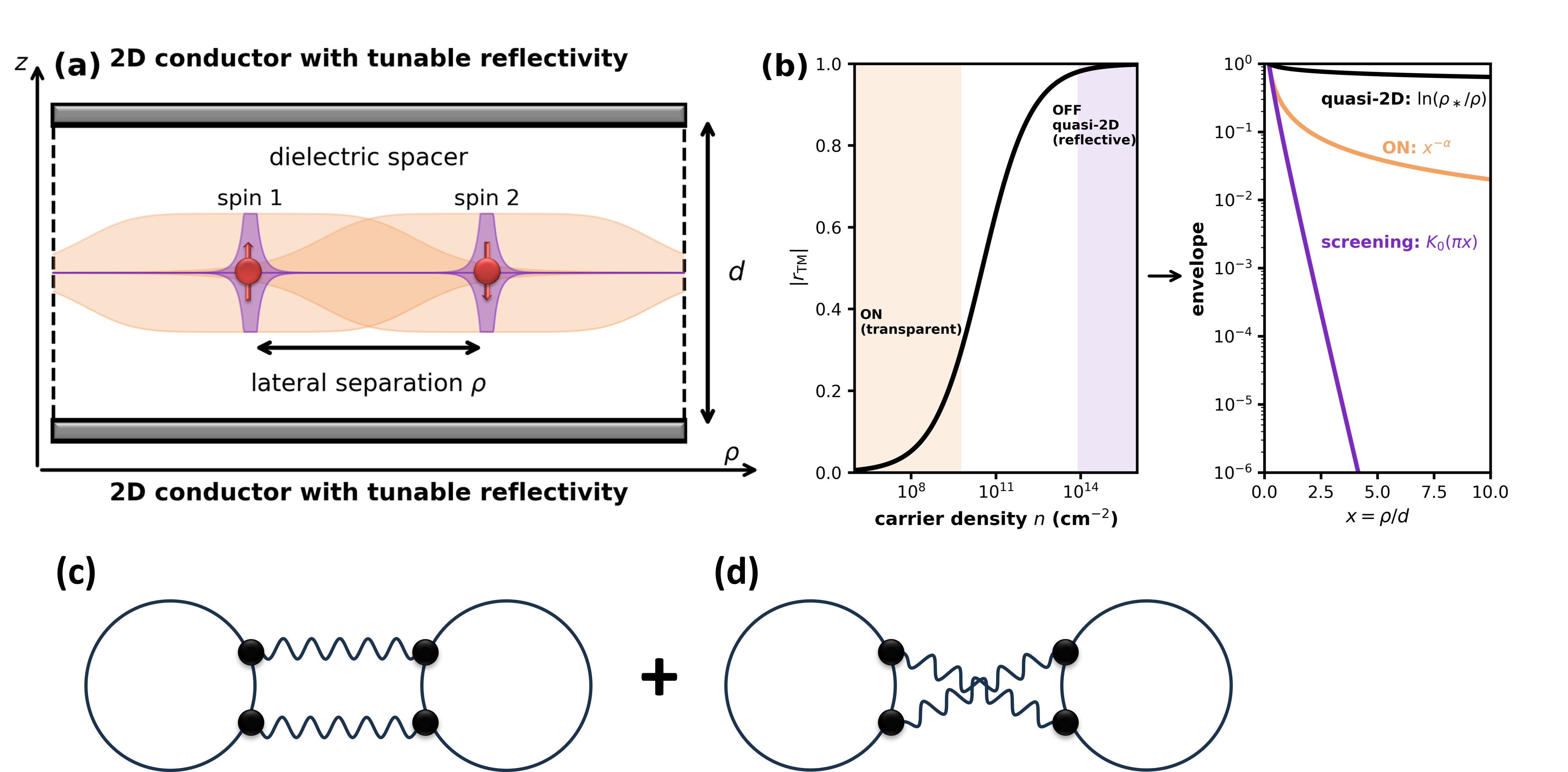}
\caption{\textbf{Universal tuning of interaction range, screening, and logarithmic antiscreening.}
\textbf{(a)} Conductor--dielectric--conductor heterostructure: a dielectric spacer of thickness \(d\)
bounded by two gate-tunable 2D conductors. Two localized objects (here illustrated as spins) reside in the
midplane and are separated by a lateral distance \(\rho\); shaded profiles schematically indicate how the
interaction can be reprogrammed by changing the electromagnetic boundary conditions.
\textbf{(b)} Gate tuning of the TM boundary response and the resulting asymptotic envelopes.
Left: representative magnitude of the static TM reflection amplitude \(|r_{\rm TM}|\) for a graphene-based
sheet model as a function of carrier density \(n\), illustrating continuous electrical tuning from a nearly
transparent regime (\(|r_{\rm TM}|\!\ll\!1\)) toward a strongly reflecting regime (\(|r_{\rm TM}|\!\to\!1\)).
Right: corresponding midplane interaction envelopes versus reduced distance \(x=\rho/d\), showing three
universal regimes: a bulk-like ON power law (\(x^{-\alpha}\)); exponential screening in the Dirichlet/PEC
parity branch (\(r_{\rm TM}\to -1\)), governed by the modified Bessel function \(K_0(\pi x)\); and a
quasi-two-dimensional logarithmic antiscreening regime in the opposite parity branch (\(r_{\rm TM}\to 1^{-}\)),
\(\mathcal D_F(\rho;0)\propto (1/d)\ln(\rho_\ast/\rho)\) for \(d\ll \rho \ll \rho_\ast\) with
\(\rho_\ast\sim d/(1-r_{\rm TM})\) (see Sec.~\ref{sec:SI_quasi-2D_log}).
\textbf{(c)} Ladder (non-crossing) two-photon exchange diagram in the linked-cluster expansion for the fluctuation-induced interaction: two local response vertices (polarizabilities \(\alpha\) for vdW/CP, or QED--DSR spin--electric vertices \(\tfrac12\,\sigma\!\cdot\!\beta\)) are connected by two cavity photon propagators \(\mathbb D_{12}(i\xi_m)\) and \(\mathbb D_{21}(i\xi_m)\), yielding the leading connected contribution \(\propto \mathrm{Tr}[\mathbb T_1\mathbb D_{12}\mathbb T_2\mathbb D_{21}]\) at Matsubara frequency \(\xi_m\).
\textbf{(d)} Crossed-ladder two-photon exchange diagram, obtained by exchanging the two photon lines. Together with panel~\textbf{c} it represents the two distinct covariant topologies of two-photon exchange (capturing all time orderings in the linked-cluster/Matsubara formulation), and underlies the compact second-order free-energy used in the main text and SI for vdW/CP and QED--DSR exchange.
}
\label{fig:hero}
\end{figure*}

Here we answer this question by developing a unified quantum-electrodynamic (QED) framework for a
conductor--dielectric--conductor heterostructure: a dielectric spacer bounded by two parallel,
gate-tunable two-dimensional conductors. Our central result is that the gate-controlled reflection
amplitudes \(r_{\mathrm{TM/TE}}(q_\perp,\omega)\) provide control not only over interaction strength but
also over interaction \emph{range and functional form}. Figure~\ref{fig:hero} summarizes the universal
switching principle and the resulting many-decade ON/OFF contrast.

\section{Green-function framework from QED}
\label{sec:qed_framework}

We formulate interaction engineering in terms of the electromagnetic Green function of the structured
environment. Starting from the QED interaction action, the interaction between two sources \(i,j\) is
\begin{equation}
S_{ij}
=\mu\,g_{\alpha\beta}\!\iint d^{4}x\,d^{4}x'\;
J_i^\alpha(x)\,\mathcal D_F(x,x')\,J_j^\beta(x'),
\label{eq:Sij_master}
\end{equation}
where \(\mathcal D_F\) is the Feynman propagator of the electromagnetic field in the presence of the
two interfaces (Lorenz gauge; details in Methods and Supplementary Information). For stationary sources,
the interaction energy follows from \(U_{ij}=-S_{ij}/T\), with \(T\) the observation time. Thus, once
\(\mathcal D_F\) (and its dyadic generalization) is known for the cavity, all interaction channels
follow.

Planar symmetry enables a Weyl (in-plane momentum) representation in cylindrical coordinates,
\begin{equation}
\mathcal D_F(\rho,z,z';\omega)
=\int_0^\infty\!\frac{dq_\perp\,q_\perp}{2\pi}\,
J_0(q_\perp\rho)\,
\widetilde{\mathcal D}_F(q_\perp;z,z';\omega),
\label{eq:Weyl_scalar}
\end{equation}
where \(\rho\) is the in-plane separation, \(q_\perp\) the in-plane wave number, and
\(\widetilde{\mathcal D}_F\) the reduced propagator in the \(z\)-direction. The reduced propagator is
obtained by summing all multiple reflections between the two sheets. Crucially, the only microscopic
inputs are the TM/TE reflection amplitudes of each sheet, \(r_{\rm TM/TE}(q_\perp,\omega)\), which are
set by the nonlocal sheet response \(\chi_{2{\rm D}}(q_\perp,\omega)\) or conductivity
\(\sigma_g(q_\perp,\omega)\) and are gate-tunable
\cite{Wunsch2006,Hwang2007,Zhu2021}.

\section{Switching power laws to exponential screening}
\label{sec:switching}

For a symmetric structure with identical interfaces at \(z=\pm d/2\) and sources at the midplane
\(z=z'=0\), the reduced Feynman propagator takes a closed form (details in SI),
\begin{equation}
\tilde\calD_{\text F}\!(0)=-\frac{1}{2Q}\left\{
\begin{aligned}
~&i+\frac{2re^{iQ d}}{2Q+ire^{iQ d}},~&\text{for }(\omega/c)^{2}\ge q_\perp^{2},\\[2pt]
~&1-\frac{2re^{-Q d}}{2Q+re^{-Q d}},~&\text{for }(\omega/c)^{2}< q_\perp^{2},
\end{aligned}
\right.
\label{eq:Dtilde_midplane}
\end{equation}
where \(Q=\sqrt{|(\omega/c)^2-q_\perp^2|}\) and \(r\) is the relevant reflection amplitude (TM/TE,
depending on the channel). Gate control enters through \(r_{\rm TM/TE}\). In particular, for TM
polarization the reflection amplitude can be written in terms of the sheet conductivity as
\begin{equation}
r_{\rm TM}(q_\perp,\omega)=
\frac{\varepsilon_2 Q_1-\varepsilon_1 Q_2
      -\dfrac{\pi e\mu c^2 q_\perp^{2}}{\omega\varepsilon_0}\,
       \sigma_g(q_\perp,\omega)}
     {\varepsilon_2 Q_1+\varepsilon_1 Q_2
      +\dfrac{\pi e\mu c^2 q_\perp^{2}}{\omega\varepsilon_0}\,
       \sigma_g(q_\perp,\omega)} ,
\label{eq:rTM_sigma_form_final}
\end{equation}
with \(Q_j=\sqrt{\varepsilon_j(\omega/c)^2-q_\perp^2}\) in medium \(j\) (and the appropriate analytic
continuation for evanescent components). The limits \(r\to 0\) (ON) and \(|r|\to 1\) (OFF) are reached by
electrostatic gating through \(\sigma_g\).
A full derivation of Eqs.~\eqref{eq:rTM_sigma_form_final} and \eqref{eq:rTE_sigma_form_final}
from Maxwell boundary conditions and a nonlocal sheet response is given in
Sec.~\ref{sec:SI_BCs} and Sec.~\ref{sec:SI_dyn_RT_ansatz} of the SI.

The universal switch emerges most transparently in the static TM sector. For sources located in the
midplane of a symmetric cavity, the static scalar Green function can be written as an image lattice
\begin{equation}
\mathcal D_F(\rho;0)=\frac{1}{4\pi\varepsilon_c}\,
\sum_{m=-\infty}^{+\infty}\frac{\bigl(r_{\rm TM}\bigr)^{|m|}}{\sqrt{\rho^2+m^2d^2}},
\qquad |r_{\rm TM}|\leq 1,
\label{eq:static_lattice}
\end{equation}
where \(r_{\rm TM}\equiv r_{\rm TM}(q_\perp,0)\) is the \emph{signed} static TM reflection amplitude of each
interface (including its phase/parity). In this convention the standard Dirichlet/PEC cavity corresponds
to \(r_{\rm TM}\to -1\) (alternating-sign images), whereas the opposite Neumann/PMC-like branch corresponds
to \(r_{\rm TM}\to +1\) (same-sign images). 
Poisson resummation reorganizes the slowly convergent image lattice into a rapidly convergent transverse-mode
(Bessel) representation \cite{Pumplin1969,Souza2015}. In particular, for the \emph{Dirichlet/PEC (screening)
branch} \(r_{\rm TM}\to -1\), only odd transverse harmonics survive at the midplane, yielding
\begin{equation}
\mathcal D_F^{\rm (PEC)}(\rho;0)=\frac{1}{4\pi\varepsilon_c}\,\frac{2}{d}
\sum_{\ell=0}^{\infty}K_0\!\Big((2\ell+1)\pi\,\rho/d\Big),
\label{eq:K0_PEC}
\end{equation}
so that the large-distance envelope is exponentially suppressed,
\begin{equation}
\mathcal D_F^{\rm (PEC)}(\rho;0)\propto \frac{e^{-\pi\rho/d}}{\sqrt{\rho/d}}
\qquad (\rho\gg d),
\label{eq:K_asymp_large}
\end{equation}
in agreement with the parallel-plate capacitor Green function \cite{Souza2015} and with the macroscopic-QED
cavity benchmark \cite{Barcellona2018}.

\begin{figure*}[t]
\centering
\includegraphics[width=\textwidth]{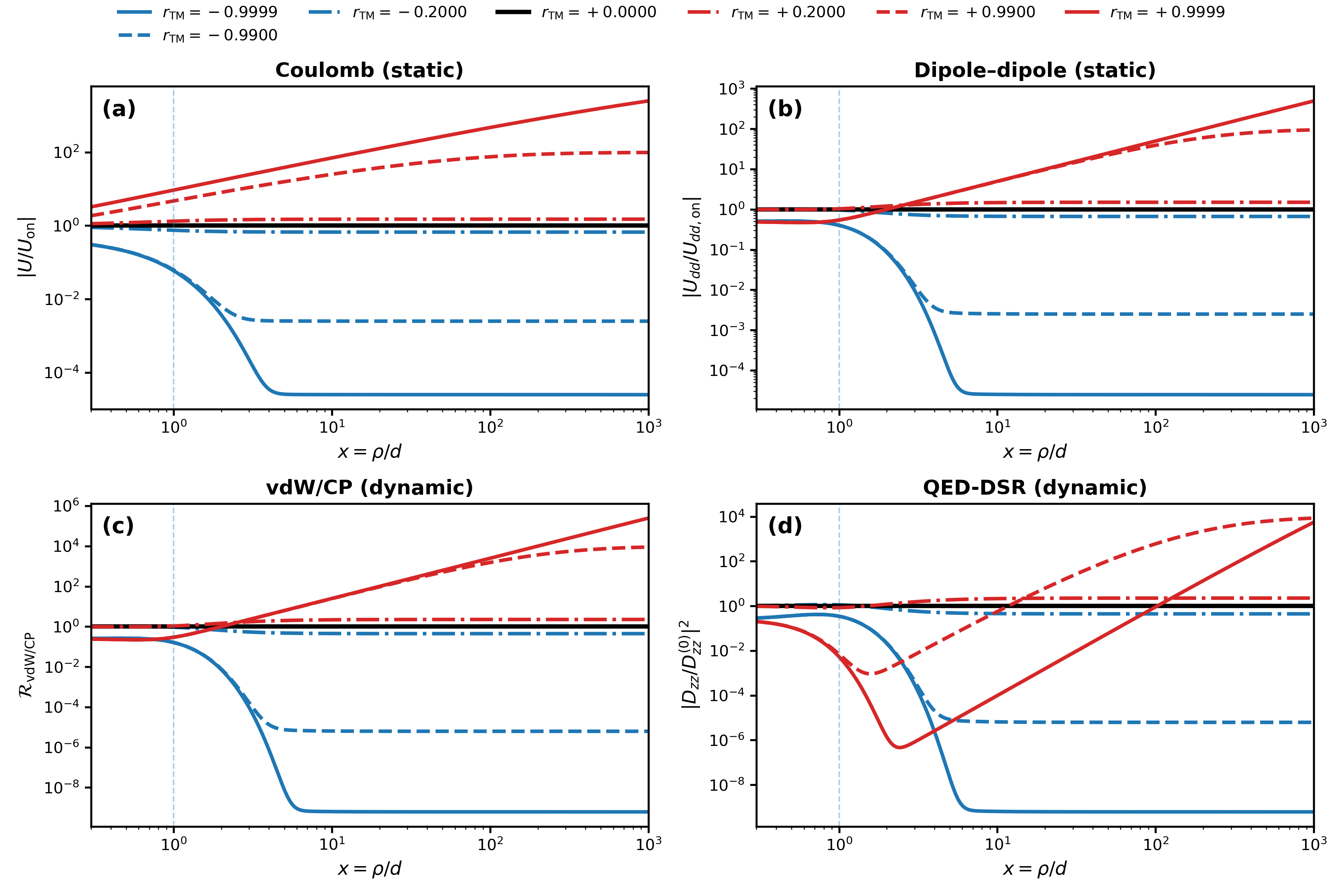}
\caption{\textbf{Tunable interaction range and strength controlled by the TM reflection amplitude.}
Normalized interactions versus reduced distance \(x=\rho/d\) for representative values of the
TM reflection amplitude \(r_{\rm TM}\) (legend), illustrating continuous tuning between bulk-like
power laws (transparent/ON limit \(r_{\rm TM}\!\approx\!0\)) and strongly modified long-range behavior
for reflective mirrors (\(|r_{\rm TM}|\!\approx\!1\)).
\textbf{(a)} Static Coulomb interaction \(U/U_{\rm on}\).
\textbf{(b)} Static electric dipole--dipole interaction \(U_{dd}/U_{dd,\rm on}\) (computed from the second
derivative of the scalar propagator; in-plane proxy).
\textbf{(c)} Fluctuation-induced vdW/CP interaction shown as a proxy proportional to the squared dipolar
Feynman propagator (capturing the characteristic \(\mathbb{D}\mathbb{D}\) dependence).
\textbf{(d)} QED-DSR-mediated spin--spin exchange shown as a proxy proportional to the squared Coulomb interaction
(capturing the characteristic \(\mathbb{D}\mathbb{D}\) dependence in the TM-dominant \(E_z\)-coupled case).
For \(|r_{\rm TM}|\ll 1\), curves are evaluated using the rapidly convergent image-lattice (power-law)
representation; for \(|r_{\rm TM}|\approx 1\), curves are evaluated using the Poisson/Schwinger
representation (Bessel-harmonic function with the exact \(\Delta S\) correction computed from the
\(\mathrm{erfc}\) / \(\mathrm{erfcx}\) form).
The sign of \(r_{\rm TM}\) selects screening (Dirichlet/PEC branch \(r_{\rm TM}\to -1\)) versus antiscreening
(Neumann/PMC-like branch \(r_{\rm TM}\to +1\)), corresponding to alternating versus same-sign image parity.
For \(r_{\rm TM}\to 1^-\), the surviving \(\ell=0\) sector produces a quasi-2D logarithmic Feynman propagator
\(\mathcal D_F(\rho;0)\propto (1/d)\ln(\rho_\ast/\rho)\) over \(d\ll\rho\ll\rho_\ast\) with
\(\rho_\ast\sim d/(1-r_{\rm TM})\), which explains the large enhancements at \(x\sim 10^2\!-\!10^3\)
for \(r_{\rm TM}=0.99\) and \(0.9999\) (Sec.~\ref{sec:SI_quasi-2D_log}).
See Sec.~\ref{sec:SI_num_Fig2} for the exact analytic expressions (including the \(\erfcx\) form)
and numerical evaluation method used to generate the curves.}
\label{fig:formalism}
\end{figure*}

By contrast, in the \emph{Neumann/PMC-like (antiscreening) branch} \(r_{\rm TM}\to +1\) the transverse \(\ell=0\)
mode survives. 
The experimentally relevant situation is \(r_{\rm TM}=1-\delta\) with \(0<\delta\ll 1\),
for which the \(\ell=0\) sector produces a wide \emph{quasi-2D logarithmic regime}:
defining \(a\equiv-\ln r_{\rm TM}\simeq \delta\) and \(\rho_\ast\sim d/a\),
one finds for
\(\,d\ll\rho\ll\rho_\ast\)
\begin{equation}
\mathcal D_F^{\rm (PMC)}(\rho;0)\;\simeq\;\frac{1}{2\pi\varepsilon_c\,d}\left[
\ln\!\Big(\frac{2d}{a\rho}\Big)-\gamma_{\rm E}\right],
\qquad (r_{\rm TM}\to 1^-),
\label{eq:log_2D_regime}
\end{equation}
i.e.\ the midplane Feynman propagator becomes effectively two-dimensional (logarithmic) up to the large crossover scale
\(\rho_\ast\sim d/(1-r_{\rm TM})\); see Sec.~\ref{sec:SI_quasi-2D_log} for a detailed derivation and its regime map.
Equations~\eqref{eq:K0_PEC} and \eqref{eq:log_2D_regime} therefore make the role of reflection parity clear:
\(r_{\rm TM}\to -1\) removes the gapless mode and yields exponential screening, whereas \(r_{\rm TM}\to 1^{-}\)
retains a gapless mode and produces a quasi-2D logarithmic antiscreening regime.
Equation~\eqref{eq:log_2D_regime} has a simple physical origin: for \(r_{\rm TM}=1-\delta\) with
\(0<\delta\ll 1\), the weights \(r_{\rm TM}^{|m|}\) decay only slowly with the image index \(|m|\),
so the discrete image lattice effectively behaves as a \emph{long (but finite) line distribution} of images
extending along \(z\) over a characteristic length \(\rho_\ast\sim d/(1-r_{\rm TM})\).
In this regime the midplane Feynman propagator can be approximated by a continuum integral
\(\sum_m r^{|m|}/\sqrt{\rho^2+m^2d^2}\to d^{-1}\!\int dz\,e^{-(a/d)|z|}/\sqrt{\rho^2+z^2}\),
so the electrostatics becomes effectively two-dimensional in the transverse plane:
integrating the 3D Coulomb interaction \(1/\sqrt{\rho^2+z^2}\) along a long direction produces the
2D Green function of the Poisson equation, \(G_{2\rm D}(\rho)\propto -\ln\rho\)
\cite{Jackson1999,Arfken_Weber2012}.
The logarithm in Eq.~\eqref{eq:log_2D_regime} is therefore the hallmark of a surviving gapless transverse mode,
and it explains the large enhancements at fixed \(x=\rho/d\) in Fig.~\ref{fig:formalism} when \(r_{\rm TM}\) is
extremely close to \(+1\). A detailed derivation and the crossover scales are given in Sec.~\ref{sec:SI_quasi-2D_log}.

For finite reflectivity \(|r_{\rm TM}|<1\) the exact resummation of the image lattice
\eqref{eq:static_lattice} is conveniently written as a Poisson/Schwinger integral,
\begin{equation}
\mathcal D_F(\rho;0)=\frac{1}{4\pi\varepsilon_c}\,\mathcal S(x;r_{\rm TM}),
\qquad x\equiv\rho/d,
\end{equation}
with
\begin{equation}
\mathcal S(x;r)=
2\sum_{\ell=0}^{\infty} I_\ell(x;r)
\end{equation}
and
\begin{equation}
I_\ell(x;r)=\frac{1}{4}\int_{0}^{\infty}\frac{ds}{s}\,e^{-s x^{2}}
\sum_{\sigma=\pm}\exp\!\Big(\frac{b_{\ell,\sigma}^{2}}{4s}\Big)\,
\erfc\!\Big(-\frac{b_{\ell,\sigma}}{2\sqrt{s}}\Big),
\label{eq:Poisson_Schwinger_erfc}
\end{equation}
where \(b_{\ell,\sigma}\equiv \ln r + i\sigma C_\ell\).
Note that the transverse harmonics are
\(C_\ell=2\pi\ell\) and the phase of \(r\) selects the parity branch: \(r\to -1\) shifts the harmonic
content to odd multiples, recovering \eqref{eq:K0_PEC}, while \(r\to +1\) retains the \(\ell=0\) mode (see Eq.~\eqref{eq:SI_K0_PMC} in the SI). For numerical stability we evaluate \eqref{eq:Poisson_Schwinger_erfc} using
the scaled complementary error function \(\erfcx(z)\equiv e^{z^{2}}\erfc(z)\), which absorbs the large
Gaussian factor and yields an exponentially convergent integrand (see Secs.~\ref{sec:SI_Poisson_resum} and \ref{sec:SI_num_Fig2} in the SI).

For static sources, the interaction energy is directly obtained from \(\mathcal D_F\).
For two charges \(e_1,e_2\) at \(z=z'=0\) separated by \(\rho\),
\begin{equation}
U^{\rm (C)}_{12}(\rho)=e_1e_2\,\mathcal D_F(\rho;z=z'=0,\omega=0).
\label{eq:Coulomb_from_DF}
\end{equation}
For two static electric dipoles \(\bd_1,\bd_2\) in the midplane,
\begin{equation}
U^{\rm (dd)}_{12}(\rho)=d_{1,i}\,\mathcal D^{\rm (stat)}_{ij}(\rho)\,d_{2,j},
\qquad
\mathcal D^{\rm (stat)}_{ij}(\rho)=\partial_i\partial_j\,\mathcal D_F(\rho;0).
\label{eq:dd_from_DF}
\end{equation}

In the Dirichlet/PEC parity branch (\(r_{\rm TM}\to -1\)), multiple reflections remove the gapless
transverse mode and the static midplane Feynman propagator is controlled by the lowest odd cavity harmonic,
producing an evanescent Bessel function and hence exponential screening at large separation
(\(x=\rho/d\gg 1\)) \cite{Pumplin1969,Souza2015,Barcellona2018}.  The corresponding asymptotic forms for the
Coulomb and static dipole--dipole interactions (power-law ON versus Bessel-\(K\) OFF, including the
near-field limits) are summarized in Table~\ref{tab:asymptotics_summary}.  In particular, the Coulomb
interaction crosses over from the bulk \(1/\rho\) law in the transparent limit to a screened
\(K_0(\pi x)\) envelope in the PEC branch, and the dipole--dipole interaction follows by differentiation,
acquiring the corresponding \(K_1(\pi x)\) screening function (Table~\ref{tab:asymptotics_summary}).

In the Neumann/PMC-like parity branch the transverse \(\ell=0\) sector survives. For \(r_{\rm TM}\) close
to \(+1\) this produces an extended quasi-2D regime in which the midplane Feynman propagator becomes logarithmic,
Eq.~\eqref{eq:log_2D_regime}. Physically, the slowly decaying weights \(r_{\rm TM}^{|m|}\) make the image
lattice behave like a long (finite-length) line distribution along \(z\); integrating the 3D Coulomb interaction
\(1/\sqrt{\rho^2+z^2}\) along this direction yields the 2D Green function \(\propto \ln\rho\).
The regime boundaries and crossover back to 3D behavior are derived in Sec.~\ref{sec:SI_quasi-2D_log},
and the component-specific nonmonotonic behavior of \(D_{zz}\) relevant for Fig.~2d is analyzed in
Sec.~\ref{sec:SI_Dzz_nonmonotonic}. In this regime,
\begin{equation}
\frac{\mathcal D_F(\rho;0)}{\mathcal D_{F,{\rm on}}(\rho;0)}
\sim 2x\Big[\ln\!\Big(\frac{2}{a x}\Big)-\gamma_{\rm E}\Big],
\label{eq:enhancement_scaling}
\end{equation}
valid for $(1\ll x\ll 1/a)$.
Here $a\equiv-\ln r_{\rm TM}\simeq 1-r_{\rm TM}$.
Note that this Feynman propagator in quasi-2D grows approximately \(\propto x\) (up to a slow logarithm), explaining the large enhancements visible
in Fig.~\ref{fig:formalism} for \(r_{\rm TM}=0.99\) and \(0.9999\) at \(x\sim 10^2\!-\!10^3\).
The static dipole--dipole interaction follows by differentiation of \(\ln\rho\) and therefore crosses over
from the 3D bulk law \(\propto 1/\rho^3\) to a quasi-2D form \(\propto 1/(d\,\rho^2)\) in the same window,
implying an analogous enhancement factor \(\sim x\) (up to logarithms) relative to the ON state.
Because vdW/CP and QED-DSR exchange involve products of Green dyadics, the quasi-2D enhancement is amplified in
those channels, scaling roughly as the square of Eq.~\eqref{eq:enhancement_scaling} in the same regime.

Equation~\eqref{eq:dd_from_DF} provides a direct construction of the static dyadic propagator as a double
derivative of the scalar propagator. Inserting the image lattice \eqref{eq:static_lattice} yields an
explicit projector form (details in Supplementary Information),
\begin{equation}
\mathcal D^{\rm (stat)}_{ij}(\rho)=
\frac{1}{4\pi\varepsilon_c}
\sum_{\ell=-\infty}^{+\infty}
\frac{\bigl(r_{\rm TM}\bigr)^{|\ell|}}{\bigl(\rho^2+\ell^2 d^2\bigr)^{3/2}}
\Bigl(3\,\hat s_{\ell,i}\hat s_{\ell,j}-\delta_{ij}\Bigr),
\label{eq:dyad_static}
\end{equation}
where \(\hat{\bs}_\ell=(\brho+\ell d\,\hat{\bz})/\sqrt{\rho^2+\ell^2 d^2}\). This compactly separates
universal geometry (the projector) from the single static material parameter \(r_{\rm TM}\).

Unlike static sources, vdW/CP interactions arise from correlated quantum fluctuations. In thermal
equilibrium, the interaction free energy between two polarizable objects can be expressed in Matsubara
form using a linked-cluster (scattering) resummation \cite{Power1982,Craig1984}:
\begin{align}
\Delta F(T)&=k_BT\sum_{m=0}^{\infty}{}'\,
\ln\det\!\Big[\mathbb I-
\mu_0\xi_m^2\alpha_1(i\xi_m)\,\mathbb D_{12}(i\xi_m)
\nn\\
&\times\mu_0\xi_m^2\alpha_2(i\xi_m)\,\mathbb D_{21}(i\xi_m)\Big].
\label{eq:vdw_logdet}
\end{align}
Here \(\mathbb D(i\xi)\) is the Euclidean electric Green dyadic of the structured environment (built from
the same multiple-reflection physics encoded in \(r_{\rm TM/TE}\)), \(\alpha_a(i\xi)\) are the polarizabilities,
and the prime denotes half-weight of the \(m=0\) term. Expanding the logarithm to second order recovers the
standard London (\(\propto \rho^{-6}\)) and Casimir--Polder (\(\propto \rho^{-7}\)) limits in bulk, while the
OFF cavity introduces a squared Bessel function through the product \(\mathbb D_{12}\mathbb D_{21}\).

For fluctuation-induced interactions, the relevant kernels are quadratic in the electromagnetic Green
dyadic: vdW/CP interactions involve \(\mathbb D_{12}\mathbb D_{21}\) in the linked-cluster free energy
\eqref{eq:vdw_logdet}, and the QED-DSR (EDSR-mediated) spin exchange \eqref{eq:Jeff_main} has the same
structure with the dipole vertex replaced by \(\beta\cdot\bsigma\).  Consequently, the cavity-induced
switching of the dyadic Green function produces an amplified ON/OFF contrast in these channels.
The asymptotic forms in the transparent limit (\(r_{\rm TM}\to 0\)), the Dirichlet/PEC screening branch
(\(r_{\rm TM}\to -1\)), and the Neumann/PMC-like logarithmic antiscreening branch (\(r_{\rm TM}\to 1^{-}\))
are collected in Table~\ref{tab:asymptotics_summary}.  In particular, the PEC branch yields a squared
Bessel function (e.g.\ \(K_0^2(\pi x)\)) and exponential suppression at \(x\gg 1\), whereas the opposite
parity branch produces a wide quasi-2D logarithmic regime that can strongly enhance the interaction at
fixed \(x\) when \(r_{\rm TM}\) is extremely close to \(+1\) (Table~\ref{tab:asymptotics_summary} and
Sec.~\ref{sec:SI_quasi-2D_log}).

The same squared-Green's function structure controls the QED-DSR spin--spin exchange interaction discussed below:
in the ON state it inherits an algebraic CP-like tail, while in the OFF state it follows
\(K_0^2(\pi x)\sim x^{-1}e^{-2\pi x}\) at large \(x\), providing a programmable range knob set by \(d\).

Electrostatic ($\omega=0$) interactions are purely TM and therefore admit complete ON/OFF control. For dynamical interactions the field decomposes into TM (radial and out-of-plane electric components) and TE (azimuthal electric component). In the quasi-electrostatic near-field, TM dominates while TE is parametrically suppressed, so both vdW/CP and QED-DSR exchange inherit the same TM-controlled Bessel-$K$ function and remain strongly tunable. If a given implementation couples primarily to azimuthal fields (TE sector), the tunability is reduced unless the interface material also provides a strongly gate-tunable TE response.

\begin{table*}[t]
\centering
\begin{threeparttable}
\caption{\textbf{Asymptotic forms of the quantum electrodynamic interactions in the three universal regimes.}
Here \(x\equiv \rho/d\). ``Transparent'' denotes \(r_{\rm TM}\to 0\) (bulk-like), the Dirichlet/PEC branch
denotes \(r_{\rm TM}\to -1\) (odd-harmonic Bessel screening), and the Neumann/PMC-like branch denotes
\(r_{\rm TM}\to 1^{-}\) (gapless mode and quasi-2D logarithmic regime).}
\label{tab:asymptotics_summary}
\small
\renewcommand{\arraystretch}{1.2}
\setlength{\tabcolsep}{4pt}
\begin{tabular}{llll}
\toprule
&
\parbox[t]{3.0cm}{\raggedright\textbf{Transparent}\\(\(r_{\rm TM}\to 0\))} &
\parbox[t]{3.5cm}{\raggedright\textbf{Dirichlet/PEC}\\(\(r_{\rm TM}\to -1\))} &
\parbox[t]{5.0cm}{\raggedright\textbf{Neumann/PMC}\\(\(r_{\rm TM}\to 1^{-}\))} \\
\midrule

\parbox[t]{3.0cm}{\raggedright\textbf{Coulomb (static)}} &
\parbox[t]{3.5cm}{\raggedright\(U_{\rm on}(\rho)\propto \rho^{-1}\)} &
\parbox[t]{4.0cm}{\raggedright\(U(\rho)\propto K_{0}(\pi x)\sim x^{-1/2}e^{-\pi x}\) for \(x\gg 1\)} &
\parbox[t]{5.0cm}{\raggedright\(\mathcal D_F(\rho;0)\simeq \dfrac{1}{2\pi\varepsilon_c d}\ln(\rho_\ast/\rho)\) for \(d\ll\rho\ll\rho_\ast\), with \(\rho_\ast\sim d/(1-r_{\rm TM})\).} \\

\parbox[t]{3.0cm}{\raggedright\textbf{Dipole--dipole (static)}} &
\parbox[t]{3.5cm}{\raggedright\(U_{dd,{\rm on}}(\rho)\propto \rho^{-3}\)} &
\parbox[t]{4.0cm}{\raggedright\(U_{dd}(\rho)\propto K_{1}(\pi x)\sim x^{-1/2}e^{-\pi x}\) for \(x\gg 1\)} &
\parbox[t]{5.0cm}{\raggedright From \(\partial_i\partial_j\ln\rho\): \(U_{dd}(\rho)\propto (d^{-1})\,\rho^{-2}\) in the quasi-2D window.} \\

\parbox[t]{3.0cm}{\raggedright\textbf{vdW/CP (dynamic)}} &
\parbox[t]{3.5cm}{\raggedright\(\Delta F_{\rm on}(\rho)\propto \rho^{-6}\) (London), \(\rho^{-7}\) (CP)} &
\parbox[t]{4.0cm}{\raggedright\(\Delta F(\rho)\propto K_0^2(\pi x)\sim x^{-1}e^{-2\pi x}\) for \(x\gg 1\)} &
\parbox[t]{5.0cm}{\raggedright In the quasi-2D window, \(\mathbb D\propto (d\,\rho^{2})^{-1}\), so \(\mathrm{Tr}[\mathbb D\!\cdot\!\mathbb D]\) and hence \(\Delta F^{(2)}\) are enhanced by \(\sim x^{2}\) relative to the bulk dyadic-squared kernel.} \\

\parbox[t]{3.0cm}{\raggedright\textbf{QED-DSR exchange (dynamic)}} &
\parbox[t]{3.5cm}{\raggedright\(J_{\rm on}(\rho)\propto \rho^{-7}\) (broad-band \(\beta\))} &
\parbox[t]{4.0cm}{\raggedright\(J(\rho)\propto K_0^2(\pi x)\sim x^{-1}e^{-2\pi x}\) for \(x\gg 1\)} &
\parbox[t]{5.0cm}{\raggedright Using \(\mathcal R_{\mathrm{QED-DSR}}=|D_{zz}/D_{zz}^{(0)}|^{2}\), the quasi-2D window gives \(D_{zz}\propto (d\,\rho^{2})^{-1}\), hence \(\mathcal R_{\mathrm{QED-DSR}}\sim x^{2}\).} \\
\bottomrule
\end{tabular}

\begin{tablenotes}[flushleft]
\footnotesize
\item The Dirichlet/PEC and Neumann/PMC-like labels refer to the effective parity branch of the static TM scalar kernel. For finite \(|r_{\rm TM}|<1\), the exact kernel is evaluated by the Poisson/Schwinger representation in stable \(\erfcx\) form and reduces to the Bessel series shown here in the perfect-mirror limits.
\end{tablenotes}
\end{threeparttable}
\end{table*}

For the static TM (electrostatic) problem, a grounded perfectly conducting mirror imposes a Dirichlet
condition on the scalar Green function, leading to the familiar alternating-sign image construction
\cite{Souza2015}. In our convention this corresponds to \(r_{\rm TM}\to -1\) in the image lattice
\eqref{eq:static_lattice}, and Poisson/Sommerfeld--Watson resummation reorganizes the resulting slowly
convergent image series into the odd-harmonic Bessel mode sum \eqref{eq:K0_PEC}, yielding exponential
screening at \(\rho\gg d\) \cite{Pumplin1969,Souza2015,Barcellona2018}. The opposite parity branch,
\(r_{\rm TM}\to +1\), corresponds to a Neumann/PMC-like (magnetic mirror) boundary for the relevant TM
scalar Feynman propagator, in which images add with the same sign and a gapless transverse mode survives; this produces a wide quasi-2D logarithmic regime for \(r_{\rm TM}\to 1^-\) (Eq.~\eqref{eq:log_2D_regime}),
leading to large enhancements at fixed \(x\) before the eventual far-field crossover; see Sec.~\ref{sec:SI_quasi-2D_log}, as illustrated in Fig.~\ref{fig:formalism}. Thus, beyond the
magnitude \(|r_{\rm TM}|\), the reflection \emph{phase/parity} provides an additional qualitative design knob
that selects screening versus antiscreening.

A particularly natural realization of \emph{fully tunable} (TM-dominated) coupling is provided by
interlayer excitons in transition-metal dichalcogenide (TMD) bilayers and twisted/moir\'e
heterobilayers.  Because the electron and hole reside in different layers, interlayer excitons carry
a \emph{permanent out-of-plane electric dipole} \(\mathbf p = e s\,\hat{\mathbf z}\), and therefore
couple primarily to the \(E_z\) component of the electromagnetic field.  In the planar TM/TE
decomposition, \(E_z\) resides in the TM sector, whereas the TE sector contains only azimuthal
in-plane fields; consequently, out-of-plane dipoles are maximally sensitive to the \emph{TM}
reflection amplitude \(r_{\rm TM}\), which is also the channel most strongly tunable by gating in
the quasi-electrostatic near field.  Moir\'e potentials in aligned or twisted TMD heterobilayers are
known to localize interlayer excitons into periodic arrays (typically with triangular-lattice
minima), producing discrete moir\'e exciton resonances and minibands
\cite{Seyler2019MoireTrapped,Tran2019MoireExcitons,Jin2019MoireExcitons,Alexeev2019Hybridized}.
Building on this moir\'e exciton platform, correlated dipolar excitonic insulators and exciton
crystals have now been realized \cite{Ma2021EI,Gu2022DipolarEI,Qi2026ExcitonCrystal}.  In our
conductor--dielectric--conductor architecture, such \(z\)-polarized dipoles inherit the full ON/OFF
switching of the TM Green function, enabling electrical control that continuously interpolates from
long-range power-law interactions to exponentially screened Bessel-\(K\) envelopes.

\section{Gate-switched spin--spin couplers}
\label{sec:spin_couplers}

We now highlight a flagship consequence for scalable quantum hardware: a gate-switched, range-programmable
spin--spin interaction mediated by electric-dipole spin resonance (QED-DSR). In EDSR, an oscillating electric
field couples to a spin degree of freedom through spin--orbit mixing and/or spin--charge hybridization,
producing an effective spin--electric response. In quantum dots and related spin qubits, EDSR is a standard
tool for electrically driven spin control \cite{Golovach2006EDSR,DuckheimLoss2006NatPhys,Nowack2007EDSR,PioroLadriere2008EDSR,Rashba2008EDSR}.
Electric-dipole spin resonance (EDSR), i.e. spin resonance driven by an AC electric field via spin–orbit coupling, was first predicted by Rashba and developed into a general framework in the classic review by Rashba and Sheka.\cite{Rashba1960,RashbaSheka1991}

We write the EDSR interaction Hamiltonian for spin \(a\) as
\begin{equation}
H_{{\rm EDSR},a}(t)=\frac{1}{2}\,\sigma_{a,i}\,\beta^{(a)}_{ij}(\omega)\,E_j(\br_a,t),
\label{eq:EDSR_vertex_main}
\end{equation}
where \(\beta_{ij}(\omega)\) is a spin--electric polarizability tensor (set by the local spin--orbit and
orbital spectrum) and \(E_j\) is the electric field at the spin location. The cavity environment shapes
the \emph{field correlator} and hence the mediated spin--spin interaction.
This compact vertex form is equivalent to the standard EDSR Hamiltonians written as an effective ac field $\mathbf{b}_1(t)\cdot\bsigma$ induced by an electric drive through spin–orbit coupling (and related mechanisms).\cite{DuckheimLoss2006NatPhys,Rashba2008EDSR}

In conventional EDSR, \(\mathbf E(t)\) is a \emph{classical} ac drive that produces single-spin rotations
\cite{Rashba2008EDSR,DuckheimLoss2006NatPhys,Golovach2006EDSR}.
Here we instead quantize both \(\hat{\mathbf E}\) and \(\hat{\boldsymbol\sigma}\) and use the EDSR vertex
as a quantum spin--field coupling: the \emph{two-spin} interaction arises from exchanging \emph{virtual photons}
between the two vertices and is fully determined by the electric Green dyadic
\(\mathbb D(\mathbf r_1,\mathbf r_2;i\xi)\).
This is a vdW/CP-type mechanism with \(\mathbf d\) replaced by \(\beta\cdot\boldsymbol\sigma\), and it is
therefore the gate-tunable cavity Green function---not an externally applied drive---that provides the
on/off and range control of the coupling.

Proceeding in Matsubara space exactly as for vdW/CP but replacing the dipole operator by the EDSR vertex,
the second-order linked-cluster contribution yields a bilinear effective spin Hamiltonian
\(\hat H_{\rm eff}=\frac{1}{2}\hat\sigma_{1,i}J_{ij}(\rho)\hat\sigma_{2,j}\) with exchange tensor
\begin{align}
J_{ij}(\rho)&=\frac{k_BT}{2}\sum_{m=0}^{\infty}{}'\,
\mu_0^2\xi_m^4\,
\Big[\beta_1(i\xi_m)\cdot \mathbb D_{12}(\rho;i\xi_m)
\nn\\
&
\cdot\beta_2(i\xi_m)\cdot \mathbb D_{21}(\rho;i\xi_m)\Big]_{ij}.
\label{eq:Jeff_main}
\end{align}
Equation~\eqref{eq:Jeff_main} clarifies the following: \emph{all} range and functional-form control
enters through \(\mathbb D(i\xi)\), which is set by the same reflection amplitudes \(r_{\rm TM/TE}\) that
control Coulomb and dipolar interactions. Thus a single gate bias on the bounding 2D conductors switches the
spin--spin interaction between ON (bulk-like) and OFF (evanescent) regimes.

In the ON state (\(r_{\rm TM/TE}\to 0\)) the dyadic reduces to the bulk Feynman propagator, and the exchange inherits
an algebraic CP-like far-field tail (for broad-band \(\beta\)),
\begin{equation}
J_{\rm ex,on}(\rho)\propto \rho^{-7}.
\label{eq:spin_on}
\end{equation}
In the OFF state (\(|r_{\rm TM/TE}|\to 1\)), the lowest harmonic yields
\(\mathbb D_{\rm off}\propto K_0(\pi x)\), so the exchange inherits
\begin{equation}
J_{\rm ex,off}(\rho)\propto K_0^2(\pi x)\sim x^{-1}e^{-2\pi x}\qquad (x\gg 1),
\label{eq:spin_off}
\end{equation}
providing an exponential range cutoff with designable length \(\ell\sim d/\pi\). Because the OFF/ON ratio
vanishes for both \(x\to 0\) and \(x\to\infty\) (Fig.~\ref{fig:hero}c), the same device can realize
(i) a strongly suppressed idle coupling (OFF) and (ii) an activated coupling (ON) at the \emph{same}
physical distance \(\rho\), with only modest differences near \(x\sim 1\).

A useful device-level figure of merit is the entangling-gate time set by the ON-state exchange scale.
For an Ising-dominated implementation,
\(
\hat H_{\rm eff}\approx \tfrac12 J_{zz}\,\hat\sigma_{1z}\hat\sigma_{2z},
\)
a controlled-phase gate (up to single-qubit rotations) requires
\(
t_{\rm CZ}\approx \pi\hbar/(2|J_{zz}|)
\).
For an XY-dominated implementation,
\(
\hat H_{\rm eff}\approx \tfrac12 J_{\perp}(\hat\sigma_{1x}\hat\sigma_{2x}+\hat\sigma_{1y}\hat\sigma_{2y}),
\)
one similarly has
\(
t_{\rm iSWAP}\approx \pi\hbar/(2|J_{\perp}|)
\).
Numerically,
\(
t_{2q}[{\rm ns}] \approx 1.03/J_{\rm eff}[\,\mu{\rm eV}\,].
\)
Thus
\(
J_{\rm eff}=0.1~\mu{\rm eV}
\)
corresponds to
\(
t_{2q}\sim 10~{\rm ns},
\)
while
\(
J_{\rm eff}=0.01~\mu{\rm eV}
\)
gives
\(
t_{2q}\sim 100~{\rm ns}.
\)
These values are intended only as representative benchmarks, since the absolute scale of
\(J_{\rm eff}\) in the present broadband DLG architecture depends on the spin-electric response
\(\beta_{ij}\), the operating distance \(\rho/d\), and the gate-controlled reflection amplitudes
\(r_{\rm TM/TE}\).

A central advantage of the present QED--DSR coupler is that its characteristic two-qubit distance is
\emph{not} fixed by wavefunction overlap.  In conventional exchange-based spin-qubit architectures,
the two-qubit interaction is generated by direct orbital overlap and therefore typically requires
nanometer-scale separations, of order \(10\!-\!100~\mathrm{nm}\), in gate-defined silicon double dots and
even more stringent placement in donor-based silicon proposals and devices
\cite{LossDiVincenzo1998,Kane1998,Veldhorst2015,Zajac2018,Madzik2021}.
By contrast, in our cavity-mediated QED--DSR scheme the relevant distance is set by the engineered electromagnetic environment rather than by direct wavefunction overlap. In the screening (Dirichlet/PEC-like) branch, the practically useful regime is
\(
1 \lesssim x \equiv \rho/d \lesssim 10,
\)
because for larger \(x\) the interaction is already strongly exponentially suppressed. Thus the coupler distance is naturally set by the spacer thickness \(d\): choosing \(d\) in the tens-to-hundreds of nanometers already moves the interaction into the submicron or micron range. For example, \(d=20~\mathrm{nm}\) gives a useful design window \(\rho\sim 20\text{--}200~\mathrm{nm}\), \(d=100~\mathrm{nm}\) gives \(\rho\sim 0.1\text{--}1~\mu\mathrm{m}\), and \(d=300~\mathrm{nm}\) gives \(\rho\sim 0.3\text{--}3~\mu\mathrm{m}\).

In the opposite, antiscreening branch, the quasi-two-dimensional logarithmic regime extends over
\(
1 \ll x \ll x_\ast, 
\) with
\(
x_\ast \sim \frac{1}{1-r_{\rm TM}},
\)
or equivalently up to the crossover distance
\(
\rho_\ast \sim \frac{d}{1-r_{\rm TM}}
\)
[cf.\ Eq.~\eqref{eq:log_2D_regime}]. Hence the accessible distance scale can become much larger than \(d\). For example, \(d=50~\mathrm{nm}\) gives \(x_\ast\sim 10^2\) and \(\rho_\ast\sim 5~\mu\mathrm{m}\) for \(r_{\rm TM}=0.99\), while \(r_{\rm TM}=0.999\) gives \(x_\ast\sim 10^3\) and \(\rho_\ast\sim 50~\mu\mathrm{m}\). Thus, unlike conventional exchange gates, whose range is fixed by nanometer-scale orbital overlap, the QED--DSR coupler makes the interaction distance a design parameter through \(d\) and, near the antiscreening branch, a gate-programmable scale through \(r_{\rm TM}\).

Scalable spin-qubit architectures require tunable two-qubit couplers that (i) can be turned off to suppress
cross-talk during single-qubit operations and idling, (ii) can be turned on reproducibly to implement
entangling gates, and (iii) do not impose strict resonant conditions or narrowband constraints. Existing
approaches to long-distance coupling rely on microwave resonators and spin-photon interfaces
\cite{Imamoglu1999,BurkardImamoglu2006,Samkharadze2018,Borjans2020,HarveyCollard2022,Dijkema2025} or on
carefully engineered electrostatic exchange networks. Our scheme provides a complementary route:
an \emph{electrically programmable interaction medium} that offers independent control of both \emph{strength}
(via \(r_{\rm TM/TE}\)) and \emph{range} (via \(d\)). This capability is particularly attractive for
large-scale layouts where wiring and connectivity constraints motivate reconfigurable, longer-range couplers.
It also naturally interfaces with van der Waals materials hosting optically addressable spins
\cite{Awschalom2018-ws,Wolfowicz2021-on}, where the conductor--dielectric--conductor environment can be
integrated as part of the heterostructure stack.

For our QED-DSR exchange interaction, the most favorable regime is a sizable out-of-plane spin--electric
polarizability, i.e.\ non-negligible tensor elements \(\beta_{iz}\) that couple the spin to the
\(E_z\) component of the field.  This choice is optimal for our conductor--dielectric--conductor cavity
because \(E_z\) resides in the TM sector, which is the polarization channel most strongly tunable by
gating in the quasi-electrostatic near field.  In conventional 3D semiconductor heterostructures,
large \(\beta_{iz}\) arises naturally when a perpendicular electric field modulates Rashba spin--orbit
coupling or the Zeeman response: representative examples include (i) EDSR in gate-defined III--V
quantum dots and double dots driven via spin--orbit mixing and engineered field gradients
\cite{Golovach2006EDSR,Nowack2007EDSR,PioroLadriere2008EDSR,Rashba2008EDSR},
(ii) disorder-enabled EDSR in spin--orbit coupled 2DEGs where an ac electric field generates an effective
internal drive field \cite{DuckheimLoss2006NatPhys},
(iii) \(g\)-tensor modulation resonance in gated heterostructures, where a time-dependent gate voltage
modulates the \(g\)-tensor and drives spin rotations \cite{Kato2003gTMR}, and
(iv) group-IV hole and acceptor spin--orbit qubits (Ge/SiGe and Si acceptors), where strong spin--orbit
physics and interface inversion asymmetry lead to pronounced electric-field control and enhanced EDSR
\cite{Watzinger2018GeHole,Hendrickx2020SingleHoleGe,Salfi2016AcceptorSOQubit}.
Closely related opportunities exist in 2D materials, where large intrinsic spin--orbit coupling and
strong vertical-field control are built into van der Waals stacks: (i) gate-defined quantum dots in
monolayer and bilayer WSe\(_2\) provide a direct 2D platform for electrically controlled spin/valley
states \cite{Davari2020WSe2QDs}, (ii) monolayer TMD spin-qubit proposals explicitly include
electric-field-induced (Bychkov--Rashba) SOC and predict efficient EDSR control \cite{Kormanyos2014PRX,BrooksBurkard2020EDSR2D},
and (iii) graphene/TMD proximity heterostructures exhibit strong induced Rashba SOC
\cite{Yang2017RashbaGrapheneTMD}, providing an all-2D route to sizable spin--electric coupling.
These platforms illustrate that the required \(\beta_{iz}\) elements are available across both
established semiconductor qubits and emerging 2D-material qubits, and they highlight clear material
pathways for implementing gate-programmable, TM-dominated QED-DSR exchange interaction in our cavity architecture.

A standard result in quantum computation is that arbitrary single-qubit rotations together with
\emph{any} entangling two-qubit interaction form a universal gate set
\cite{NielsenChuang2000,Barenco1995}.  In spin-qubit platforms, single-qubit control is routinely
implemented via electrically driven spin resonance (EDSR) enabled by Rashba-type spin--orbit coupling,
while two-qubit entangling gates are commonly generated by pulsing a controllable exchange interaction
\cite{LossDiVincenzo1998,Golovach2006EDSR,Rashba2008EDSR}.  In this context, our gate-programmable,
range-tunable QED-DSR exchange interaction \(\hat H_{\rm eff}=\tfrac12\,\hat\sigma_{1,i}J_{ij}(\rho)\hat\sigma_{2,j}\)
provides a fully electrical, switchable entangling resource for two-spin gates; combined with standard
single-spin EDSR control, it supplies the essential ingredients for universal spin-based quantum
computing.  The key new capability is that the \emph{same} gate knob controls both the strength and
the spatial range of the two-qubit coupling, enabling low-crosstalk idling (OFF) and activated gates
(ON) in scalable layouts.

\section{Design rules and outlook}
\label{sec:outlook}

Our results establish two-dimensional conductor--dielectric--conductor heterostructures as a universal
platform for \emph{gate control of interaction range and functional form}. The platform is defined by two
independent knobs: a \emph{geometric range knob} set by the spacer thickness \(d\), which controls the
lowest evanescent harmonic and hence the exponential decay length \(\ell\sim d/\pi\), and a
\emph{material/electrostatic knob} set by the reflection amplitudes \(r_{\rm TM/TE}\), which control the
interpolation between bulk power laws (ON) and evanescent Bessel functions (OFF). Because the formalism is
expressed in terms of reflection data \(r_{\rm TM/TE}(q_\perp,\omega)\), it is material-agnostic and applies
to any gate-tunable 2D conductor with a well-defined nonlocal response.

For scalable spin-qubit hardware, the QED-DSR exchange \eqref{eq:Jeff_main} provides an electrically programmable
two-qubit coupler whose range and strength can be tuned independently. A practical design workflow is:
(i) choose \(d\) to set the target interaction range \(\ell\sim d/\pi\) at the device separation \(\rho\);
(ii) use gate bias to set \(r_{\rm TM/TE}\) and thus the on/off ratio at fixed \(x=\rho/d\); and (iii)
shape gate pulses to switch between idle (OFF) and gate (ON) operation windows. This interaction-medium
approach complements resonator-based couplers by providing broadband, geometry-set range control without
requiring narrowband cavity resonance.

Several extensions are natural. Lossy and temperature-dependent sheet responses modify \(r_{\rm TM/TE}\) and
broaden the Bessel envelope, but the ON/OFF switching principle remains intact. Patterned or anisotropic
conductors can make \(r_{\rm TM/TE}\) angle dependent, enabling directional interaction engineering.
Nonequilibrium settings can be treated by replacing Matsubara dyadics with Keldysh Green functions.
These directions preserve the central design rule: once reflection data are known (or engineered),
interactions follow by universal transforms, providing reconfigurable interaction landscapes for quantum
matter and quantum hardware.

A timely opportunity is to use programmable screening/antiscreening as a \emph{diagnostic knob} for the
pairing mechanism in magic-angle twisted bilayer graphene (MATBG), where the origin of superconductivity
remains actively debated. Experiments have already demonstrated that proximate screening layers can tune
correlation strength and superconductivity in MATBG \cite{Liu2021Screening} and can even suppress
superconductivity in nearby moir\'e graphene by Coulomb screening \cite{Barrier2024Screening}.
On the theory side, both phonon-mediated scenarios \cite{Wu2018PhononTBG,Chen2024StrongEPhMATBG} and
purely electronic/plasmon-assisted scenarios \cite{Cea2021CoulombPhononsTBG,Peng2024ScreeningGatePlasmon}
have been proposed. Our interaction-medium architecture provides a direct route to vary the effective
long-range Coulomb interaction \(\mathcal V(q,i\xi)\propto 1/\varepsilon(q,i\xi)\) over a wide range by tuning
\(r_{\rm TM}(q,i\xi)\) and the spacer length scale \(d\). In the simplest interpretation, if superconductivity
is insensitive to strong Coulomb suppression it points toward predominantly phonon-driven pairing, whereas
strong sensitivity to screening supports an electronic (Coulomb/plasmon) mechanism. Conversely, if an
electronic mechanism dominates, engineering the antiscreening branch and the relevant retarded spectral
weight raises the tantalizing possibility of substantially enhancing the pairing scale;  this idea can be
tested quantitatively by tracking \(T_c\) as a function of programmable screening.

This work has been supported by the Office of Naval Research (ONR) through the U.S. Naval Research Laboratory (NRL). M. N. L. acknowledges support by the Air Force Office of Scientific Research (AFOSR) under award no. FA9550-23-1-0472.

\bibliography{bibliography}

\end{document}